\documentclass[12pt]{article} 

\usepackage{latexsym} 	
\usepackage{amssymb}  	
\usepackage{amsmath}  	
\usepackage{amsbsy}
\usepackage{epsfig}     

\allowdisplaybreaks

\newcommand{\Tr}{\mbox{Tr}}

\newcommand{\C}{{\cal C}}
\newcommand{\bra}{\langle}
\newcommand{\ket}{\rangle}
\newcommand{\sgn}{\mbox{sgn}}
\newcommand{\half}{\frac{1}{2}}

\newcommand{\be}{\begin{equation}}
\newcommand{\ee}{\end{equation}}
\newcommand{\bea}{\begin{eqnarray}}
\newcommand{\eea}{\end{eqnarray}}
\newcommand{\bean}{\begin{eqnarray*}}
\newcommand{\eean}{\end{eqnarray*}}
\newcommand{\nn}{\nonumber}
\newcommand{\hm}{\hspace*{-0.6cm}}

\newcommand{\NNLOp}{{N$^2$LO$^\prime$} }

\renewcommand{\theequation}{\arabic{section}.\arabic{equation}}

\begin{document}

\title{
\vskip -120pt
{\begin{normalsize}
\mbox{} \hfill SWAT 06/463\\
\mbox{} \hfill DAMTP-2006-28\\
\mbox{} \hfill hep-th/0604156 \\
\vskip  10pt
\end{normalsize}}
{\bf\Large
Nonequilibrium dynamics in the $O(N)$ model to next-to-next-to-leading 
order in the $1/N$ expansion 
}
\author{
\addtocounter{footnote}{2}
Gert Aarts$^{a}$\thanks{email: g.aarts@swan.ac.uk}
 {} and
Anders Tranberg$^{b,c}$\thanks{email: a.tranberg@damtp.cam.ac.uk}
 \\ {} \\
{}$^a${\em\normalsize Department of Physics, University of Wales Swansea}
\\
{\em\normalsize Singleton Park, Swansea, SA2 8PP, United Kingdom}
\\ {} \\
 {}$^b${\em\normalsize Department of Physics \& Astronomy, University of 
Sussex} \\
   {\em\normalsize Falmer, Brighton, BN1 9QH, United Kingdom}
\\ {} \\
 {}$^c${\em\normalsize DAMTP, University of Cambridge} \\
   {\em\normalsize Wilberforce Road, Cambridge, CB3 0WA, United Kingdom}
}
}
\date{April 21, 2006}
\maketitle
\begin{abstract}

Nonequilibrium dynamics in quantum field theory has been studied 
extensively using truncations of the 2PI effective action. Both $1/N$ and 
loop expansions beyond leading order show remarkable improvement when 
compared to mean-field approximations. However, in truncations used so 
far, only the leading-order parts of the self energy responsible for 
memory loss, damping and equilibration are included, which makes it 
difficult to discuss convergence systematically. For that reason 
we derive the real and causal evolution equations for an $O(N)$ model to 
next-to-next-to-leading order in the 2PI-$1/N$ expansion. Due to the 
appearance of internal vertices the resulting equations appear intractable 
for a full-fledged $3+1$ dimensional field theory. Instead, we solve the 
closely related three-loop approximation in the auxiliary-field formalism 
numerically in $0+1$ dimensions (quantum mechanics) and compare to 
previous approximations and the exact numerical solution of the 
Schr\"odinger equation.

\end{abstract}
                                                                                
\newpage
                                                                                
                                                                                
                                                                     
\section{Introduction}
\setcounter{equation}{0}

The two-particle irreducible (2PI) effective action formalism has proven 
very powerful for out-of-equilibrium quantum field theory over a wide 
range of applications \cite{Berges:2004yj}. Since it necessarily employs 
an expansion and truncation of the effective action, one should be 
concerned with how well these expansions converge.\footnote{Of course, 
perturbative or $1/N$ expansions in quantum mechanics and field theory are 
usually not convergent but instead asymptotic, see e.g.\ ref.\ 
\cite{zinnjustin} for discussions concerning the $O(N)$ model.}

An extensive study has been made of effective mean-field approximations 
(Gaussian, leading order (LO) $1/N$ or Hartree approximations) which 
amounts to including a self-consistent, time-dependent effective mass in 
the dynamics of the one- and two-point functions \cite{LOapp}. However, 
since such an approximation does not include scattering, phenomena such as 
damping, memory loss and equilibration cannot be described properly. This 
can be traced back to the existence of (infinitely many) conserved charges 
in the mean-field dynamics and the presence of a nonthermal fixed point 
\cite{Bettencourt:1997nf,Aarts:2000wi}.\footnote{Some scattering is 
present if one allows for inhomogeneous mean fields 
\cite{Aarts:1999zn,Salle:2000hd,Bettencourt:2001xg}.}
 Going beyond the mean-field approximation, by either employing the 
weak-coupling or the $1/N$ expansion to next-to-leading order (NLO), 
effective memory loss, damping and equilibration are 
present.\footnote{\label{footnote1}See e.g.\ refs.\ 
\cite{Berges:2000ur,Aarts:2001qa,Juchem:2003bi,Arrizabalaga:2005tf} for 
the loop expansion, refs.\ 
\cite{Berges:2001fi,Mihaila:2000sr,Aarts:2001yn,Blagoev:2001ze,Aarts:2002dj,
Berges:2002cz,Cooper:2002qd,Arrizabalaga:2004iw} for the 2PI-$1/N$ 
expansion to NLO, and ref.\ \cite{Berges:2002wr} for 
2PI dynamics with fermions.}
 In particular it has been found that at late times the system evolves 
towards a quantum equilibrium state, characterized by suitably defined 
field occupation numbers approaching the familiar Bose-Einstein or 
Fermi-Dirac distribution functions. The self-consistently determined 
propagators have a time-dependent mass and width. It is found therefore 
that NLO approximations improve dramatically upon mean-field 
approximations and qualitatively reproduce the dynamics expected on 
physical grounds from the full, untruncated system.

The natural question to ask is whether a truncation at NLO also gives 
quantitatively correct results. This issue has been investigated in several 
ways, usually by comparing the NLO results to other available 
approximations. For example, when the coupling is small, normal 
perturbation theory should be applicable for e.g.\ estimates of damping 
rates. Indeed, the results in ref.\ \cite{Arrizabalaga:2005tf} suggest 
that the perturbative result is reproduced within a factor of two for 
small coupling, where the difference is due to the effect of including 
self-consistent infinite resummations in the 2PI approach. However, this 
amounts to a test of perturbation theory rather than a verification of the 
2PI formalism. Similarly, it is well-known \cite{Calzetta:1986cq} how to 
derive on-shell kinetic (Boltzmann) equations from truncations of the 2PI 
effective action and one can compare the self-consistent 2PI dynamics with 
dynamics from kinetic theory 
\cite{Juchem:2003bi,Lindner:2005kv,Berges:2005md}. However, 
this again serves more as a test of kinetic theory than of the 2PI 
truncation. 
  Another possibility, relevant for the dynamics at very late times, is to 
study transport coefficients from the 2PI effective action, which gives
insight into what scattering processes are included 
\cite{Calzetta:1999ps,Aarts:2003bk}.

 When the exact, untruncated evolution is accessible, one may carry out a 
direct comparison. This option is available in quantum mechanics, where 
the dynamics from the 2PI effective action can be benchmarked against the 
numerical solution of the Schr\"odinger equation 
\cite{Mihaila:2000ib}.
 Finally, perhaps the most detailed comparison has been made within 
classical statistical field theory, where direct numerical simulations are 
straightforward \cite{Aarts:2000wi}.
 Moreover, the 2PI formalism can be easily applied to classical dynamics 
of an initial nonequilibrium ensemble 
\cite{Aarts:2001yn,Blagoev:2001ze,Arrizabalaga:2004iw}.\footnote{In 
quantum field theory, stochastic quantization techniques have recently 
yielded the first real-time nonequilibrium lattice simulation results 
\cite{Berges:2005yt}. When developed further, this would offer the 
possibility for direct tests as well.}
  In ref.\ \cite{Aarts:2001yn} it was found in the context of the $1+1$ 
dimensional $O(N)$ model that the nonperturbative classical evolution, 
obtained numerically, is well reproduced by the 2PI-$1/N$ expansion at NLO 
for $N$ larger than about $10$, providing direct support for the use of 
the 2PI-$1/N$ approximation at NLO.
	
\begin{figure}[t]
\centerline{\psfig{figure=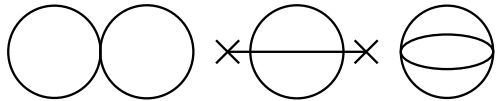,width=6.5cm}}
\caption{
Weak-coupling expansion of the 2PI effective action in the broken phase of 
a scalar $\lambda\phi^4$ theory \cite{Arrizabalaga:2005tf}.
}
\label{figweak}
\end{figure}

  Ideally, within a given approximation scheme, the validity of a 
truncation is estimated by extending the method to the next order in the 
expansion and test for effective convergence. Due to the absence of 
scattering in mean-field approximations, it is essential to go to NLO and 
compare NLO with approximations that go further.
 In the coupling expansion, a first attempt to do this was made in ref.\ 
\cite{Arrizabalaga:2005tf}. Here it was noted that in the broken phase of 
a $\lambda\phi^{4}$ theory two diagrams contribute at 
$\mathcal{O}(\lambda)$: the Hartree and the background field dependent 
diagram, shown in fig.\ \ref{figweak} on the left. Perturbatively neither 
of these diagrams leads to (on-shell) damping, but after the 
self-consistent 2PI resummation the second one does. This approximation 
can then be extended by inclusion of the basketball diagram, shown in 
fig.\ \ref{figweak} on the right. From a comparison between the dynamical 
evolution in the two truncations, it was found that equilibration times 
differ, but only by a factor less than two, see ref.\ 
\cite{Arrizabalaga:2005tf} for more details.
  Both truncations are at the level of complexity of the 2PI-$1/N$ 
expansion to NLO and are therefore numerically tractable. 
  Technically, the common feature of these truncations, and in fact of all 
truncations treated so far, is the absence of internal vertices in self 
energies.
 This is important from a numerical point of view, which is complicated 
due to the presence of memory integrals in the evolution equations. 
Additional vertices require extra memory integrals to be carried out, 
which is numerically expensive.

In this paper we extend the analysis and include for the first time self 
energies with internal vertices, using the framework of the 2PI-$1/N$ 
expansion to next-to-next-to-leading order (N$^2$LO) in the symmetric 
phase of the $O(N)$ model. This opens up the possibility to make a 
quantitative comparison between evolution at NLO and N$^2$LO in the $1/N$ 
expansion.
 The N$^2$LO effective equations of motion contain additional (nested) 
space-time integrals for each time step, when compared to NLO. It is 
beyond the scope of this paper to solve those equations in full field 
theory. For the sake of illustration we instead specialise to quantum 
mechanics which allows us to test the consistency of the equations, the 
conservation of energy and effective convergence of the expansion. We 
stress that the dynamics of quantum mechanics is of course very different 
from field theory.
  Given sufficient computer power the evolution equations can readily be 
implemented in field theory.

 The paper is organized as follows. In the next section we briefly review 
the 2PI-$1/N$ expansion in the $O(N)$ model, following closely the 
discussion and notation of ref.\ \cite{Aarts:2002dj}. In section 
\ref{causalequations} we present the dynamical equations to N$^2$LO and 
give explicitly (part of) the statistical and spectral self energies, 
which are much more involved than at NLO due to the presence of internal 
vertices. Results from the numerical implementation for the $0+1$ 
dimensional case are shown in section \ref{quantummechanics}, while the 
outlook is given in section \ref{outlook}. In three appendices we collect 
technicalities related to the standard loop expansion, multi-loop contour 
integrals and the numerical solution of the Schr\"odinger equation.


\section{2PI-$1/N$ expansion\label{2PIformalism}}
\setcounter{equation}{0}

Throughout the paper we consider an $N$ component scalar field in the 
$O(N)$ symmetric phase ($\langle \phi_{a}\rangle=0$). The action is
 \be
\label{eqS}
S = - \int_\mathcal{C} d^4x \left[ \half \partial_\mu 
\phi_a\partial^\mu \phi_a  
+ \half m^2 \phi_a\phi_a +\frac{\lambda}{4!N}(\phi_a\phi_a)^2
\right],
\ee
 with $a=1,\ldots,N$. Doubled indices are summed over. As appropriate to 
an out-of-equilibrium treatment, the fields are defined along the Keldysh 
contour $\mathcal{C}$ in the complex-time plane, see appendix 
\ref{appcontourintegrals}. The 2PI effective 
action depends on the full two-point function $G$ of the theory and can be 
parametrised as \cite{Cornwall:1974vz} 
 \be 
\Gamma[G] = \frac{i}{2} \Tr\ln G^{-1}
          + \frac{i}{2} \Tr\, G_0^{-1} \left(G-G_0\right)
          + \Gamma_2[G], 
\label{2PIaction}
\ee
 where $G_0^{-1}$ denotes the free inverse propagator. Variation of the 
effective action with respect to $G$ results in the Schwinger-Dyson 
equation for the two-point function,
$G^{-1} = G_{0}^{-1} - \Sigma$, which after multiplication with 
$G$ reads 
\be
\label{SchwingerDyson}
-\left[\square_x + m^2 \right] G_{ab}(x,y) =
i \int_z \Sigma_{ac}(x,z) G_{cb}(z,y) + i \delta_{ab}
\delta_{\C}(x-y).
\ee
Here we used the short-hand notation
\be
\int_z = \int_{\cal C} d^4z.
\ee
The self energy is given by
\be
\label{eqSigma2pi}
\Sigma_{ab} =2i\frac{\delta\Gamma_2[G]}{\delta G_{ab}}.
\ee
 The full four-point function is represented by the nonlocal term on the 
RHS of 
eq.\ (\ref{SchwingerDyson}). For future reference we note here that
\be
 \label{eqfourp}
   \frac{\lambda}{6N} \bra \phi_a^2(x) \phi_c^2(x)\ket  =  
   i \int_\C d^4z\, \Sigma_{ac}(x,z) G_{ca}(z,x),
 \ee
 which can be verified using e.g.\ the Heisenberg equations of motion.

To implement the $1/N$ expansion efficiently, it is convenient to use the 
auxiliary-field formalism 
\cite{Coleman:jh,Abbott:1975bn,Mihaila:2000sr,Aarts:2002dj}. 
The action then reads
\be
 S[\phi,\chi] = - \int_x \left[ \half \partial_\mu 
\phi_a\partial^\mu \phi_a  + \half m^2 \phi_a\phi_a 
 -\frac{3N}{2\lambda}\chi^2 + \frac{1}{2}\chi\phi_a\phi_a  \right].
\ee
 Integrating out $\chi$ yields the original action (\ref{eqS}). The 2PI
effective action is now written in terms of the one-point 
function $\bar\chi =\bra\chi(x)\ket$ and the two-point functions
\be
G_{ab}(x,y) =  \bra T_\C \phi_a(x) \phi_b(y)\ket, 
\;\;\;\;\;\;\;\;
D(x,y) =  \bra T_\C \chi(x) \chi(y)\ket - 
\bra\chi(x)\ket\bra\chi(y)\ket,
\ee
and reads
\bea
\nn
\Gamma[G,D, \bar\chi] = &&\hm
 S[0,\bar\chi]+ \frac{i}{2} \Tr\ln G^{-1} + \frac{i}{2} \Tr\, G_0^{-1} 
\left(G-G_0\right)
\\ && \hm 
+ \frac{i}{2} \Tr\ln D^{-1} + \frac{i}{2} \Tr\, D_0^{-1} \left(D-D_0\right)
          + \Gamma_2[G,D].
\label{eqGamma}
\eea
 Since we take $\bra\phi_a\ket=0$, there is no mixing between the $\phi$ 
and $\chi$ propagators \cite{Mihaila:2000sr,Aarts:2002dj}. The free 
inverse propagators read
\be
G_{0,ab}^{-1}(x,y) = 
i\left[ \square_x +m^2+\bar\chi(x)\right]\delta_{ab}\delta_\C(x-y),
\;\;\;\;\;\;\;\;
D_0^{-1}(x,y)=\frac{3N}{i\lambda}\delta_\C(x-y). 
\ee
The evolution equations for the propagators $G$ and $D$ and the one-point 
function $\bar\chi$ are obtained by extremizing (\ref{eqGamma}) and
read
\bea
 -\left[ \square_x + m^2 +\bar\chi(x) \right] G_{ab} (x,y) = &&\hm
  i\int_z \Sigma_{ac}(x,z) G_{cb}(z,y)
 + i\delta_{ab} \delta_\C(x-y),
\nn \\
 \frac{3N}{\lambda} D(x,y) = &&\hm 
 i\int_z \Pi (x,z) D(z,y) + i\delta_\C(x-y),
\eea
and
\be
 \bar\chi(x) = \frac{\lambda}{6N}G_{cc}(x,x).
\ee
\begin{figure}[t]
\centerline{\psfig{figure=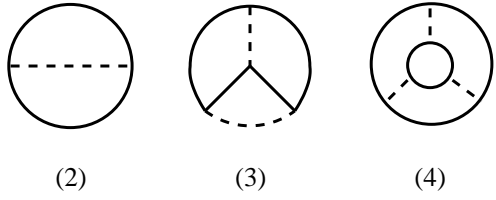,width=6.5cm}}
\caption{
NLO (2 loops) and N$^2$LO (3 and 4 loops) contributions in the 2PI-$1/N$ 
expansion. The scalar propagator $G$ is denoted with the full line and the 
auxiliary-field propagator $D$ with the dashed line.
}
\label{figNLONNLO}
\end{figure}
The self energies are defined by
\be
\label{eqSigmaPi}
\Sigma_{ab} =2i\frac{\delta\Gamma_2[G,D]}{\delta G_{ab}},
\;\;\;\;\;\;\;\;
\;\;\;\;\;\;\;\;
\Pi =2i\frac{\delta\Gamma_2[G,D]}{\delta D}.
\ee
It is convenient to separate the local part of $D$ \cite{Aarts:2002dj}
and write
 \be
 \label{eqDhat}
 D(x,y) = \frac{\lambda}{3N}\left[
i\delta_{\C}(x-y)+ \hat D(x,y) \right],
\ee
such that $\hat D$ is determined from
\be
 \hat D(x,y) = -\frac{\lambda}{3N} \Pi(x,y)
 + \frac{i\lambda}{3N} \int_z  \Pi(x,z) \hat D(z,y).
\ee

\begin{figure}[t]
\centerline{\psfig{figure=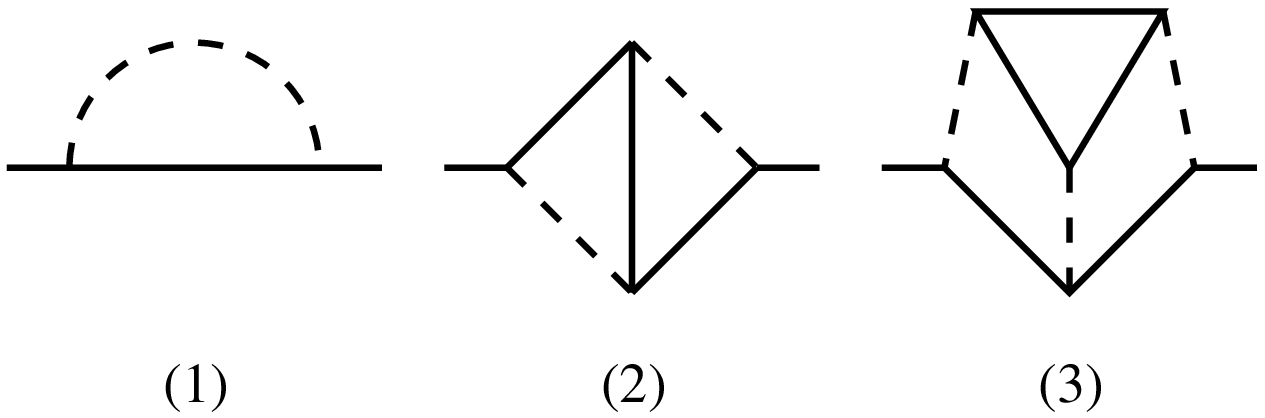,width=8.5cm}}
\caption{Self energy $\Sigma$ at NLO (1 loop) and N$^2$LO (2 and 3 loops).
}
\label{figsigma}
\vspace*{0.5cm}
\centerline{\psfig{figure=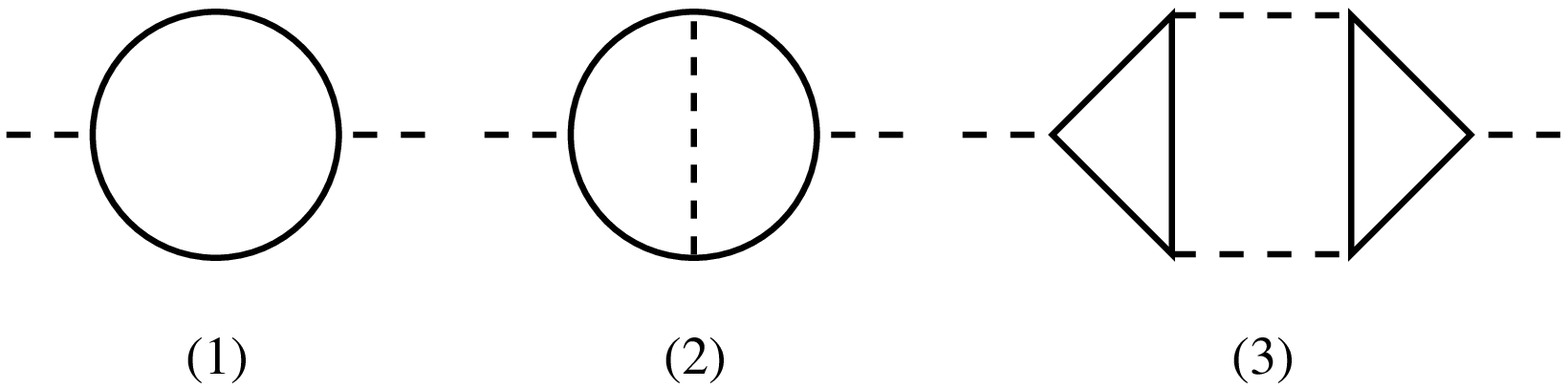,width=10.5cm}}
\caption{Auxiliary-field self energy $\Pi$ at NLO (1 loop) and N$^2$LO (2 
and 3 loops).
}
\label{figpi}
\end{figure}

We now continue with the 2PI-$1/N$ expansion, in which there is one 
diagram at NLO and two diagrams at N$^2$LO, see fig.\ \ref{figNLONNLO}. 
For a detailed powercounting discussion we refer to ref.\ 
\cite{Aarts:2002dj}. It suffices here to say that a closed scalar 
propagator $G\sim N$ and the auxiliary-field propagator $D\sim 1/N$. It is 
then easy to see that diagram $(2)\sim 1$ and diagrams $(3)$ and $(4) \sim 
1/N$. The expressions are
 \bea
\Gamma_2^{\rm NLO}[G,D] = &&\hm \frac{i}{4}
\int_{xy} G_{ab}^2(x,y)D(x,y), \\
\nn
\Gamma_2^{\rm NNLO(3)}[G,D] = &&\hm -\frac{i}{8}
\int_{xyzw} G_{ab}(x,y) G_{bc}(y,z) G_{cd}(z,w) G_{da}(w,x)
D(x,z)D(y,w), \\
\nn
\Gamma_2^{\rm NNLO(4)}[G,D] = &&\hm \frac{i}{12}
\int_{xyzx'y'z'} 
G_{ab}(x,y) G_{bc}(y,z) G_{ca}(z,x) \\
\nn
&&\hm \times
G_{a'b'}(x',y') G_{b'c'}(y',z') G_{c'a'}(z',x')
D(x,x')D(y,y')D(z,z').
\eea
 The corresponding self energies are shown in figs.\ \ref{figsigma} and 
\ref{figpi}. 

\begin{figure}[t]
\centerline{\psfig{figure=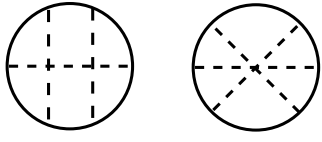,height=1.8cm}}
\caption{Loop expansion in the auxiliary-field formalism: additional 
four-loop diagrams, of order $1/N^2$ (N$^3$LO).  
}
\label{fig4loops}
\end{figure}

We continue with the $O(N)$ symmetric case, such that $G_{ab}(x,y)= 
G(x,y)\delta_{ab}$ and $\Sigma_{ab}(x,y)= \Sigma(x,y)\delta_{ab}$. For 
notational simplicity we label the self energies according to the number 
of loops, e.g.\ $\Sigma^{({\ell})}$, where ${\ell}=1,2,3$. We stress that 
the 2PI-$1/N$ expansion does not coincide with loop expansion in the 
auxiliary-field formalism. For instance, there are two more four-loop 
diagrams, see fig.\ \ref{fig4loops}, which only contribute at N$^3$LO.  
The connection with the standard loop expansion is discussed further in 
appendix \ref{apploopexpansion}.

At NLO the self energies read 
\bea
\Sigma^{(1)}(x,y) = &&\hm -G(x,y)D(x,y), \\ 
\Pi^{(1)}(x,y) = &&\hm -\frac{N}{2} G^2(x,y),
\eea
and at N$^2$LO
\bea
\label{eqsigma2}
\Sigma^{(2)}(x,y)= &&\hm 
\int_{zw} G(x,w) G(w,z) G(z,y) D(x,z)D(w,y), \\
\nn
\Sigma^{(3)}(x,y)= &&\hm -N\int_{zx'y'z'} 
\!\!\!\!\!\!\!\!
G(x,z) G(z,y) G(x',y') G(y',z') G(z',x')
D(x,x')D(y,y')D(z,z'), \\
\label{eqpi2}
\Pi^{(2)}(x,y)= &&\hm \frac{N}{2}
\int_{zw} G(x,w) G(w,y) G(x,z) G(z,y) D(z,w), \\
\nn
\Pi^{(3)}(x,y)= &&\hm -\frac{N^2}{2} \int_{zz'ww'} 
\!\!\!\!\!\!\!\!\!\!\!
G(x,z) G(z,w) G(w,x) G(y,z') G(z',w') G(w',y)
D(w,w') D(z,z').
\eea

\section{Causal equations\label{causalequations}}
\setcounter{equation}{0}

To bring the evolution equations in a form that can be solved numerically, 
the contour propagators and self energies are written in terms of 
statistical ($F$) and spectral ($\rho$) components \cite{Aarts:2001qa},
 \be
\label{eqG}
G(x,y) = F(x,y)-\frac{i}{2} \sgn_{\cal C}(x,y)\rho(x,y),
\ee
 and similar for $\hat D$, $\Sigma$ and $\Pi$. Here $\sgn_{\cal C}(x,y)= 
\Theta_{\C}(x^0-y^0)- \Theta_{\C}(y^0-x^0)$ is the sign-function along the 
contour, see appendix \ref{appcontourintegrals}.
The explicitly causal equations then read \cite{Aarts:2002dj}
\bea
\left[\square_x + M^2(x) \right] \rho(x,y) =  &&\hm  
 -\int_{y^0}^{x^0} dz \, \Sigma_\rho(x,z)\rho(z,y), 
 \label{rhoeq}\\  
\nn
 \left[\square_x + M^2(x) \right] F(x,y)  = &&\hm  
 -\int_0^{x^0} dz \, \Sigma_\rho(x,z) F(z,y)
 +\int_0^{y^0} dz \, \Sigma_F(x,z)\rho(z,y),
 \label{Feq}
\eea
where
\be
\label{hartreemass1}
M^2(x) = m^2 + \lambda\frac{N+2}{6N} F(x,x),
\ee
and
\bea
 \frac{3N}{\lambda} \hat D_\rho(x,y) = &&\hm  
 - \Pi_\rho(x,y)
 + \int_{y^0}^{x^0} dz\,\Pi_\rho(x,z) \hat D_\rho(z,y),
 \label{Drhoeq}\\ 
\nn
\frac{3N}{\lambda}\hat D_F(x,y) = &&\hm  - \Pi_F(x,y)
+ \int_{0}^{x^0} dz\, \Pi_\rho(x,z) \hat D_F(z,y)
- \int_{0}^{y^0} dz\, \Pi_F(x,z) \hat D_\rho(z,y).
 \label{DFeq}
\eea
We use here the notation
\be
\int_{y^0}^{x^0} dz = \int_{y^0}^{x^0} dz^0 \int d^3z. 
\ee
Equations of motion derived from truncations of the 2PI effective 
action conserve the following energy functional (cf.\ eq.\ 
(\ref{eqfourp})),
\bea
 E/N = &&\hm \int d^3x\, \half \left[  \partial_{x^0}\partial_{y^0} 
+  \partial_{x^i}\partial_{y^i} +  m^2 
+ \frac{\lambda}{12} \frac{N+2}{N} F(x,x) \right] F(x,y)\Big|_{x=y} 
\nn \\ &&\hm
+
\frac{1}{4} \int d^3x \int_0^{x^0} dz^0 \int d^3z\, \left[ 
\Sigma_\rho(x,z)F(z,x)
- \Sigma_F(x,z)\rho(z,x) \right].
\label{eqenergy}
\eea
The statistical and spectral self energies at NLO can be found in ref.\ 
\cite{Aarts:2002dj} and read
\bea
 \Sigma_F^{(1)}(x,y) = &&\hm -\frac{\lambda}{3N} \left[
 F(x,y)\hat D_F(x,y) - \frac{1}{4} \rho(x,y)\hat D_\rho(x,y) \right],
\nn \\
 \Sigma_\rho^{(1)}(x,y) =&&\hm -\frac{\lambda}{3N} \left[
 \rho(x,y)\hat D_F(x,y) + F(x,y)\hat D_\rho(x,y) \right],
\nn \\
\Pi_F^{(1)}(x,y) = &&\hm -\frac{N}{2} \left[ F(x,y) F(x,y) -
 \frac{1}{4}\rho(x,y) \rho(x,y) \right],
\nn \\
 \Pi_{\rho}^{(1)}(x,y) = &&\hm - NF(x,y)\rho(x,y).
\label{Prhoeq}
\eea

We now come to the causal self energies at N$^2$LO. These self energies 
have internal vertices and therefore require further contour integrals. We 
start by discussing $\Sigma^{(2)}(x,y)$ in some detail. Since we have 
separated the local part of $D$, we first insert eq.\ (\ref{eqDhat}) into 
eq.\ (\ref{eqsigma2}). This yields
 \bea
\Sigma^{(2)}(x,y) = &&\hm
\int_{zw} D(x,z) G(x,w) G(y,z) D(y,w)  G(z,w)
\nn \\ =  &&\hm
- g^2 G^3(x,y) 
\nn \\ &&\hm
+ i g^2 \int_z \left[
\hat D(x,z) G(x,y) G^2(y,z)
+ G^2(x,z) G(x,y) \hat D(y,z) \right]
\nn \\ &&\hm
+ g^2 \int_{zw} \hat D(x,z) G(x,w) G(y,z) \hat D(y,w) G(z,w),
\eea
 where 
\be
 g=\frac{\lambda}{3N}.
\ee
 This is naturally organised according to the number of $\hat{D}$ 
propagators and we use the notation $\Sigma^{(2,n)}$ for the contribution 
with $n$ $\hat{D}$'s. The term without $\hat D$ propagators is the N$^2$LO 
contribution to the setting-sun diagram and reads
 \bea 
 \Sigma^{(2,0)}_F(x,y) = &&\hm - g^2 
\left[ F^2(x,y)-\frac{3}{4}\rho^2(x,y) \right]F(x,y), 
\nn \\ 
 \Sigma^{(2,0)}_\rho(x,y) = &&\hm - g^2 \left[ 3F^2(x,y) - 
 \frac{1}{4}\rho^2(x,y) \right]\rho(x,y). 
\label{S1a} 
\eea
 The other terms quickly become rather lengthy, so we first discuss the 
general structure. Every line in the self energy can be either a 
statistical ($F$) or a spectral ($\rho$) function. With $L$ lines, this 
gives a maximum of $2^L$ possibilities. However, in order to find a 
nonzero result, every internal vertex needs at least one sgn-function 
coming from a $\rho$-type line ending on it (see appendix 
\ref{appcontourintegrals}). This implies that a diagram with $V$ internal 
vertices should have at least $V$ $\rho$-type lines.
 This reduces the maximal number of terms in the expressions for the 
causal 
diagrams to 
\be
 \# \,\,\mbox{distinct contributions} \leq  \sum_{l_\rho=V}^{L} 
\binom{L}{l_\rho} 
\leq 2^L,
\ee
 where $l_\rho$ is the number of $\rho$-type lines. 
  The actual number of nonzero contributions is in fact slightly less, 
since some contributions vanish after performing the contour integrals, 
due to the appearance of internal vertices without $\rho$-type lines (even 
though $l_\rho \geq V$).
 Finally, the number of terms that have to be independently 
evaluated is further reduced due to the fact that statistical (spectral) 
self energies are explicitly even (odd) under interchange of $x$ and $y$.
 We also note that if the contour self energy is proportional to $i$ (such 
as $\Sigma^{(2,1)}$), expressions with an odd number of $\rho$-type 
lines contribute to $\Sigma_F$ and expressions with an even number 
of $\rho$-type lines to $\Sigma_\rho$.
 If the contour self energy is proportional to $1$ (such as 
$\Sigma^{(2,2)}$), this is reversed. 
 
 Using the contour integration rules summarized in appendix
\ref{appcontourintegrals}, we find explicitly
\bea
\frac{1}{g^{2}}
\nn &&\hm 
\Sigma^{(2,1)}_F(x,y) = 
\\ &&\hm
+ \int_0^x dz\, \left[ 2 F(x,z)\rho(x,z) \hat D_F(y,z)
+ \hat D_\rho(x,z) \left(F^2(y,z)-\frac{1}{4}\rho^2(y,z)
\right) \right] F(x,y)
\nn \\ &&\hm
+ \int_0^y dz\, \left[ 2\hat D_F(x,z) F(y,z)\rho(y,z)
+  \left(F^2(x,z)-\frac{1}{4}\rho^2(x,z)
\right) \hat D_\rho(y,z) \right] F(x,y)
\nn \\ &&\hm
+  \int_y^x dz\, \half  \left[ \hat D_\rho(x,z) F(y,z)\rho(y,z)
+ F(x,z)\rho(x,z) \hat D_\rho(y,z) \right] \rho(x,y),
\label{S1b}
\eea
and
\bea
\frac{1}{g^{2}}
&&\hm \nn 
\Sigma^{(2,1)}_\rho(x,y) = 
\\ &&\hm
+ \int_0^x dz\, \left[ 2 F(x,z)\rho(x,z) \hat D_F(y,z)
+ \hat D_\rho(x,z) \left(F^2(y,z)-\frac{1}{4}\rho^2(y,z)
\right) \right] \rho(x,y)
\nn \\ &&\hm
+ \int_0^y dz\, \left[ 2\hat D_F(x,z) F(y,z)\rho(y,z)
+ \left(F^2(x,z)-\frac{1}{4}\rho^2(x,z) \right) \hat D_\rho(y,z) \right] 
\rho(x,y)
\nn \\ &&\hm
-  \int_y^x dz\, 2 \left[ \hat D_\rho(x,z) F(y,z)\rho(y,z)
 + F(x,z)\rho(x,z) \hat D_\rho(y,z) \right] F(x,y),
\eea
and for the self energy with two $\hat D$ propagators 
\bea
\frac{1}{g^{2}}
\Sigma^{(2,2)}_F(x,y) =  
\nn &&\hm 
 - \int_0^x dz\int_0^x dw\, 
 \rho(x,z) \hat D_\rho(x,w) \hat D_F(y,z) F(y,w) F(z,w) 
\\ &&\hm \nn
 - \int_0^y dz\int_0^y dw\, 
 \hat D_F(x,z) F(x,w) \rho(y,z) \hat D_\rho(y,w) F(z,w) 
\nn \\ &&\hm
 - \int_0^x dz\int_0^z dw\, \Big[ 
 \rho(x,z) \hat D_F(x,w) \hat D_F(y,z) F(y,w) 
\nn \\ &&\hm \hspace*{2.6cm}
 +
 \hat D_F(x,z) F(x,w) F(y,z) \hat D_\rho(y,w) 
 \Big] \rho(z,w) 
\nn \\ &&\hm 
 - \int_0^y dz\int_0^z dw\, \Big[ 
 \hat D_F(x,z) F(x,w) \rho(y,z) \hat D_F(y,w)  
\nn \\ &&\hm \hspace*{2.6cm}
 +
 F(x,z) \hat D_\rho(x,w) \hat D_F(y,z) F(y,w) 
 \Big] \rho(z,w) 
\nn \\ &&\hm 
 - \int_0^x dz\int_0^y dw\, \Big[
  \hat D_\rho(x,z) F(x,w) F(y,z) \hat D_\rho(y,w) 
\nn \\ &&\hm \hspace*{2.6cm}
 +
 \rho(x,z) \hat D_F(x,w) \hat D_F(y,z) \rho(y,w) 
 \Big]  F(z,w)
\nn \\ &&\hm
 - \int_0^x dz\int_z^y dw\, \frac{1}{4} \Big[
 \hat D_\rho(x,z) F(x,w) \rho(y,z) \hat D_\rho(y,w)
\nn \\ &&\hm \hspace*{2.6cm}
 +
 \rho(x,z) \hat D_F(x,w) \hat D_\rho(y,z) \rho(y,w)
 \Big] \rho(z,w) 
\nn \\ &&\hm 
 - \int_0^y dz\int_z^x dw\, \frac{1}{4}\Big[
 \hat D_\rho(x,z) \rho(x,w) \rho(y,z) \hat D_F(y,w) 
\nn \\ &&\hm \hspace*{2.6cm}
 +
 \rho(x,z) \hat D_\rho(x,w) \hat D_\rho(y,z) F(y,w) 
 \Big] \rho(z,w)
\nn \\ &&\hm 
 + \int_y^x dz\int_y^x dw\, \frac{1}{4}  
 \hat D_\rho(x,z) \rho(x,w) \rho(y,z) \hat D_\rho(y,w) F(z,w),
\label{Sf1c}
\eea
and
\bea
\frac{1}{g^{2}}
\Sigma^{(2,2)}_\rho(x,y) 
= 
&&\hm
 + \int_y^x dz\int_0^x dw\, \Big[ 
 \hat D_\rho(x,z) \rho(x,w) \rho(y,z) \hat D_F(y,w) 
\nn \\ &&\hm \hspace*{1.2cm}
+
 \rho(x,z) \hat D_\rho(x,w) \hat D_\rho(y,z) F(y,w) 
\Big] F(z,w)
\nn \\ &&\hm 
 + \int_y^x dz\int_0^y dw\, \Big[
 \hat D_\rho(x,z) F(x,w) \rho(y,z) \hat D_\rho(y,w) 
\nn \\ &&\hm \hspace*{1.2cm}
+
 \rho(x,z) \hat D_F(x,w) \hat D_\rho(y,z) \rho(y,w) 
\Big] F(z,w)
\nn \\ &&\hm
 + \int_y^x dz\int_0^z dw\, \Big[
 \hat D_\rho(x,z) F(x,w) \rho(y,z) \hat D_F(y,w) 
\nn \\ &&\hm \hspace*{1.2cm}
+ 
 \rho(x,z) \hat D_F(x,w) \hat D_\rho(y,z) F(y,w) 
\Big] \rho(z,w)
\nn \\ &&\hm 
 + \int_y^x dz\int_y^z dw\, \half \Big[ 
 \hat D_\rho(x,z) F(x,w) F(y,z) \hat D_\rho(y,w) 
\nn \\ &&\hm \hspace*{1.2cm}
+
 \rho(x,z) \hat D_F(x,w) \hat D_F(y,z) \rho(y,w) 
\Big] \rho(z,w)
\nn \\ &&\hm 
 + \int_y^x dz\int_x^z dw\,\half \Big[ 
 F(x,z) \hat D_\rho(x,w)  \hat D_\rho(y,z) F(y,w) 
\nn \\ &&\hm \hspace*{1.2cm}
+
 \hat D_F(x,z) \rho(x,w)  \rho(y,z) \hat D_F(y,w) 
 \Big] \rho(z,w) 
\nn \\ &&\hm
 - \int_y^x dz\int_0^y dw\, \frac{1}{4} \Big[ 
 \hat D_\rho(x,z) \rho(x,w) \rho(y,z) \hat D_\rho(y,w) 
\nn \\ &&\hm \hspace*{1.2cm}
+
 \rho(x,z) \hat D_\rho(x,w) \hat D_\rho(y,z) \rho(y,w) 
\Big] \rho(z,w).
\label{Sr1c}
\eea
In a few terms we used the symmetry of the integrand to make some minor
simplifications.

For the auxiliary-field self energy we proceed in the same manner and 
find, at two loops,
\bea
\Pi^{(2)}(x,y)= &&\hm \frac{N}{2}
\int_{zw} G(x,z) G(x,w) G(y,z) G(y,w) D(z,w)
  \\ \nn
= &&\hm \frac{i\lambda}{6} 
\int_{z} G^2(x,z) G^2(y,z) 
+
\frac{\lambda}{6}
\int_{zw} G(x,z) G(x,w) G(y,z) G(y,w) \hat D(z,w).
\eea
Denoting these diagrams again as $\Pi^{(2,n)}$, where $n$ denotes the 
number of $\hat D$ propagators, we find with $n=0$, 
\bea
\label{PI1a}
\Pi^{(2,0)}_F(x,y) = &&\hm +  \frac{\lambda}{3}  \int_0^x dz\, 
F(x,z) \rho(x,z) \left( F^2(y,z) -\frac{1}{4} \rho^2(y,z) \right)
\nn \\ &&\hm 
+ \frac{\lambda}{3} \int_0^y dz\, 
\left( F^2(x,z) -\frac{1}{4} \rho^2(x,z) \right) F(y,z) \rho(y,z), 
\nn \\
\Pi^{(2,0)}_\rho(x,y) = &&\hm -\frac{2\lambda}{3} \int_y^x dz\, 
F(x,z) \rho(x,z) F(y,z) \rho(y,z),
\eea
and with one $\hat D$ propagator
\bea
\label{PIf1b}
\frac{6}{\lambda}\Pi^{(2,1)}_F(x,y) = 
&&\hm  
 - \int_0^x dz\int_0^x dw\, 
 \rho(x,z) \rho(x,w) F(y,z) F(y,w) \hat D_F(z,w) 
\nn \\ &&\hm 
 - \int_0^y dz\int_0^y dw\, 
 F(x,z) F(x,w) \rho(y,z) \rho(y,w) \hat D_F(z,w) 
\nn \\ &&\hm
 - \int_0^x dz\int_0^z dw\, 2
 \rho(x,z) F(x,w) F(y,z) F(y,w) \hat D_\rho(z,w) 
\nn \\ &&\hm 
 - \int_0^y dz\int_0^z dw\, 2
 F(x,z) F(x,w)  \rho(y,z) F(y,w) \hat D_\rho(z,w) 
\nn \\ &&\hm
 - \int_0^x dz\int_0^y dw\, 2
 \rho(x,z) F(x,w) F(y,z) \rho(y,w) \hat D_F(z,w) 
\nn \\ &&\hm 
 + \int_y^x dz\int_y^x dw\, \frac{1}{4} 
 \rho(x,z) \rho(x,w) \rho(y,z) \rho(y,w) \hat D_F(z,w) 
\nn \\ &&\hm
 - \int_0^x dz\int_z^y dw\, \half  
 \rho(x,z) F(x,w) \rho(y,z) \rho(y,w) \hat D_\rho(z,w) 
\nn \\ &&\hm 
 - \int_0^y dz\int_z^x dw\, \half 
 \rho(x,z) \rho(x,w) \rho(y,z) F(y,w) \hat D_\rho(z,w),
\label{PIr1b}
\eea
and
\bea
\frac{6}{\lambda}\Pi^{(2,1)}_\rho(x,y) = 
&&\hm  
 + \int_y^x dz\int_0^x dw\, 2
 \rho(x,z) \rho(x,w) \rho(y,z) F(y,w) \hat D_F(z,w)
\nn \\ &&\hm 
 + \int_y^x dz\int_0^y dw\, 2
 \rho(x,z) F(x,w) \rho(y,z) \rho(y,w) \hat D_F(z,w)
\nn \\ &&\hm
 + \int_y^x dz\int_y^z dw\, 
 \rho(x,z) F(x,w) F(y,z) \rho(y,w) \hat D_\rho(z,w)
\nn \\ &&\hm 
 + \int_y^x dz\int_x^z dw\, 
 F(x,z) \rho(x,w) \rho(y,z) F(y,w) \hat D_\rho(z,w)
\nn \\ &&\hm
 + \int_y^x dz\int_0^z dw\, 2
 \rho(x,z) F(x,w) \rho(y,z) F(y,w) \hat D_\rho(z,w)
\nn \\ &&\hm 
 - \int_y^x dz\int_0^y dw\, \half
 \rho(x,z) \rho(x,w) \rho(y,z) \rho(y,w) \hat D_\rho(z,w).
\eea
 Since the $\Pi$ self energies have more internal symmetry than the 
$\Sigma$ self energies, the corresponding expressions are slightly 
shorter. We also emphasize that for nonequilibrium quantum fields
statistical ($F$) and spectral ($\rho$) components are independent 
\cite{Aarts:2001qa}:  there are therefore no (obvious) cancellations 
between the various terms above.
 
 We now briefly discuss the requirements for a numerical solution. The 
evolution equations for the two-point functions (\ref{rhoeq}) and 
(\ref{Drhoeq}) can be solved numerically by 
discretisation on a space-time lattice. For this it is necessary to 
perform the space-time integrals on the RHS of those 
equations. This has been discussed extensively in references cited in 
footnote \ref{footnote1}.
 At NLO, the self energies (\ref{Prhoeq}) are simple products of $F$, 
$\rho$, $\hat D_F$ and $\hat D_\rho$. At N$^2$LO however, the self 
energies themselves require space-time integrals to be performed: the 
two-loop self energy presented above contains up to two internal vertices, 
and the three-loop self energy will contain up to four internal vertices. 
These lead to nested loops over time, dramatically increasing the 
necessary CPU time.
  Because of the obvious numerical effort required, we have not attempted 
to solve the full N$^2$LO approximation. Instead we concentrate in the 
next section on the three-loop approximation (two-loop self energy) in the 
$0+1$ dimensional case.

 The statistical and spectral self energies corresponding to the 
three-loop self energies $\Sigma^{(3)}$ and $\Pi^{(3)}$ in eqs.\ 
(\ref{eqsigma2}) and (\ref{eqpi2}) are very lengthy. For instance, for 
$\Sigma_F^{(3,3)}$ one has to consider 44 distinct nonzero diagrams which 
differ in the way the statistical and spectral functions appear on the 8 
internal lines. Because we are not including these diagrams in the 
numerical analysis below, we refrain from giving the explicit expressions. 
Using the contour integrals listed in appendix \ref{appcontourintegrals}
it is straightforward, albeit cumbersome, to work them out.


\section{Quantum dynamics
\label{quantummechanics}}
\setcounter{equation}{0}

Since a numerical solution of the full $3+1$ dimensional field theory 
seems intractable, we now restrict our attention to quantum mechanics, 
which in this context is equivalent to 0+1 dimensional field theory. 
We consider the Hamiltonian
\be
 \mathcal{H} = \frac{1}{2}p_{a}p_{a} + \frac{1}{2}m^2q_{a}q_{a} + 
\frac{\lambda}{4!N}\left(q_{a}q_{a}\right)^{2}, 
\ee
and consider $O(N)$ symmetric Gaussian initial conditions, parametrized
by the normalized Gaussian density matrix \cite{Cooper:1996ii},
\be
\bra q'| \hat\rho(\eta,\xi,\sigma)|q\ket = 
\frac{1}{(2\pi\xi^2)^{N/2}}\exp\left[ 
-\frac{\sigma^2+1}{8\xi^2}(q_a'^2+q_a^2) + 
i\frac{\eta}{2\xi}(q_a'^2-q_a^2) + 
\frac{\sigma^2-1}{4\xi^2} q_aq_a' \right].
\label{eqdens}
\ee
Recall that we consider $\bra q_a\ket=\bra p_a\ket=0$. 
When $\sigma=1$, this reduces to the density matrix of a pure state,
\be
\hat\rho(\eta,\xi,0) = |\Psi_0\ket\bra\Psi_0|,
\ee
with
\be
\Psi_0(q) = \bra q|\Psi_0\ket = 
\frac{1}{(2\pi\xi^2)^{N/4}}\exp\left[ 
- \left( \frac{1}{4\xi^2} + i\frac{\eta}{2\xi}\right) q_a^2
\right].
\ee
In this case, the Schr\"odinger equation 
 \be
i\frac{\partial}{\partial t}\Psi(q_a,t) = {\cal H}\Psi(q_a,t),
\ee
can be solved numerically (see appendix \ref{appsch}), which allows for a 
comparison with the untruncated evolution 
\cite{Mihaila:2000ib}.\footnote{In ref.\ \cite{Mihaila:2000ib} only 
density matrices with $\eta=0$ (and $\sigma=1$) were considered.}

The expressions from the previous sections remain unchanged, provided that 
all reference to space indices and integrals are dropped, e.g.,
\bea
 F_{ab}(x,y) &&\hm \to F_{ab}(t,t') = \delta_{ab}F(t,t') = 
\half \bra q_a(t)q_b(t') + q_b(t') q_a(t)\ket, \\
 \rho_{ab}(x,y) &&\hm \to \rho_{ab}(t,t') = \delta_{ab}\rho(t,t') = 
 i\bra [q_a(t),q_b(t')]\ket.
\eea
  The density matrix (\ref{eqdens}) yields the following initial 
conditions for the statistical function
 \bea
\nn
F(t,t')\Big|_{t=t'=0} = &&\hm \xi^2,\\
\partial_t F(t,t')\Big|_{t=t'=0} = &&\hm \xi\eta,\\
\partial_t\partial_{t'} F(t,t')\Big|_{t=t'=0} = &&\hm
\left(\eta^2+\frac{\sigma^2}{4\xi^2}\right).
\nn 
\eea
As always, the spectral function satisfies
\be
\rho(t,t) =  0,
\;\;\;\;\;\;\;\;\;\;\;\;
\partial_t \rho(t,t')\Big|_{t=t'} = 1.
\ee
The initial conditions for $\hat D_F$ and $\hat D_\rho$ are determined by 
eq.\ (\ref{Drhoeq}). The (conserved) energy takes the value
\be
\bra {\cal H}\ket/N = \half \eta^2 +\frac{\sigma^2}{8\xi^2} +\half m^2 
\xi^2 +\frac{\lambda}{4!} \frac{N+2}{N}\xi^4.
\ee

 We solve the evolution equations (\ref{rhoeq}) and (\ref{Drhoeq}) 
numerically for various $\lambda$ and $N$ using a simple leap-frog 
algorithm and a standard discretization of the time integrals. 
 We set $m=1$ throughout. For a step size of $dt=0.01$ the entire memory 
kernel fits on a 1GB CPU, for runs until $t=20$ (2000 timesteps). Using 
$dt=0.001$ leads to indistinguishable results. A run which includes the 
first N$^2$LO diagram takes about 2 days on a single 3GHz machine, 
although presumably this can be improved somewhat. The CPU-time grows as 
the number of timesteps to the 4th power. A run at NLO is roughly 400 
times faster. Including another nested integral (i.e.\ including the 
second N$^{2}$LO diagram) is expected to give at least another factor of 
400, making such an extension very challenging indeed.

\begin{figure}[t]
\centerline{\psfig{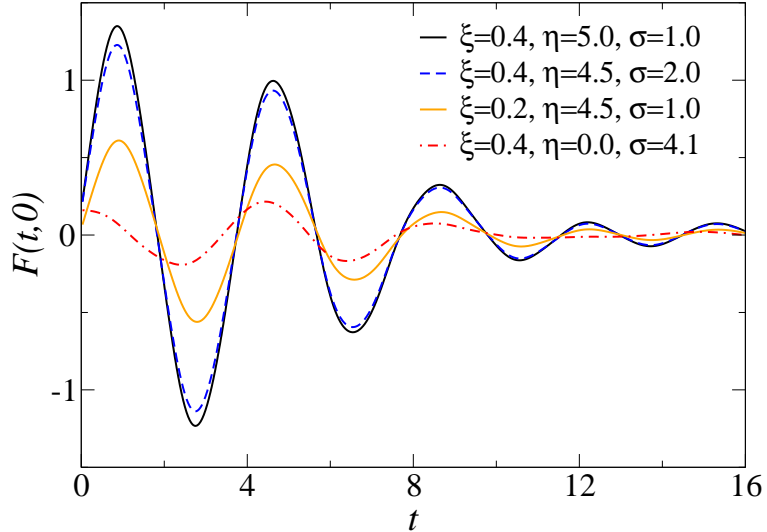}}
\caption{Unequal time correlator $F(t,0)$ at NLO for various 
choices of initial states with the same energy ($N=8$, $\lambda=1$). 
 }
\label{initcond2}
\end{figure}

In order to study the effect of the different initial conditions, we show 
in fig.\ \ref{initcond2} the time evolution at NLO for $N=8$ and 
$\lambda=1$, for various choices of initial states with the same energy. 
The early evolution, $0<t<0.5$, can be understood from the uncoupled case 
($\lambda=0$), in which the dynamics is readily solved, and the two-point 
functions are
 \bea
\nn
F(t,t') = &&\hm 
\xi^2\cos(mt)\cos(mt')
+ \frac{1}{m^2}\left( \eta^2+\frac{\sigma^2}{4\xi^2}\right) \sin(mt) \sin(mt')
+ \frac{\xi\eta}{m} \sin\left[m(t+t')\right], \\
\rho(t,t') = &&\hm \frac{\sin\left[m(t-t')\right]}{m}.
\eea
  The subsequent evolution is of course very different from the free case, 
in which damping is absent.

\begin{figure}
\centerline{\psfig{figure=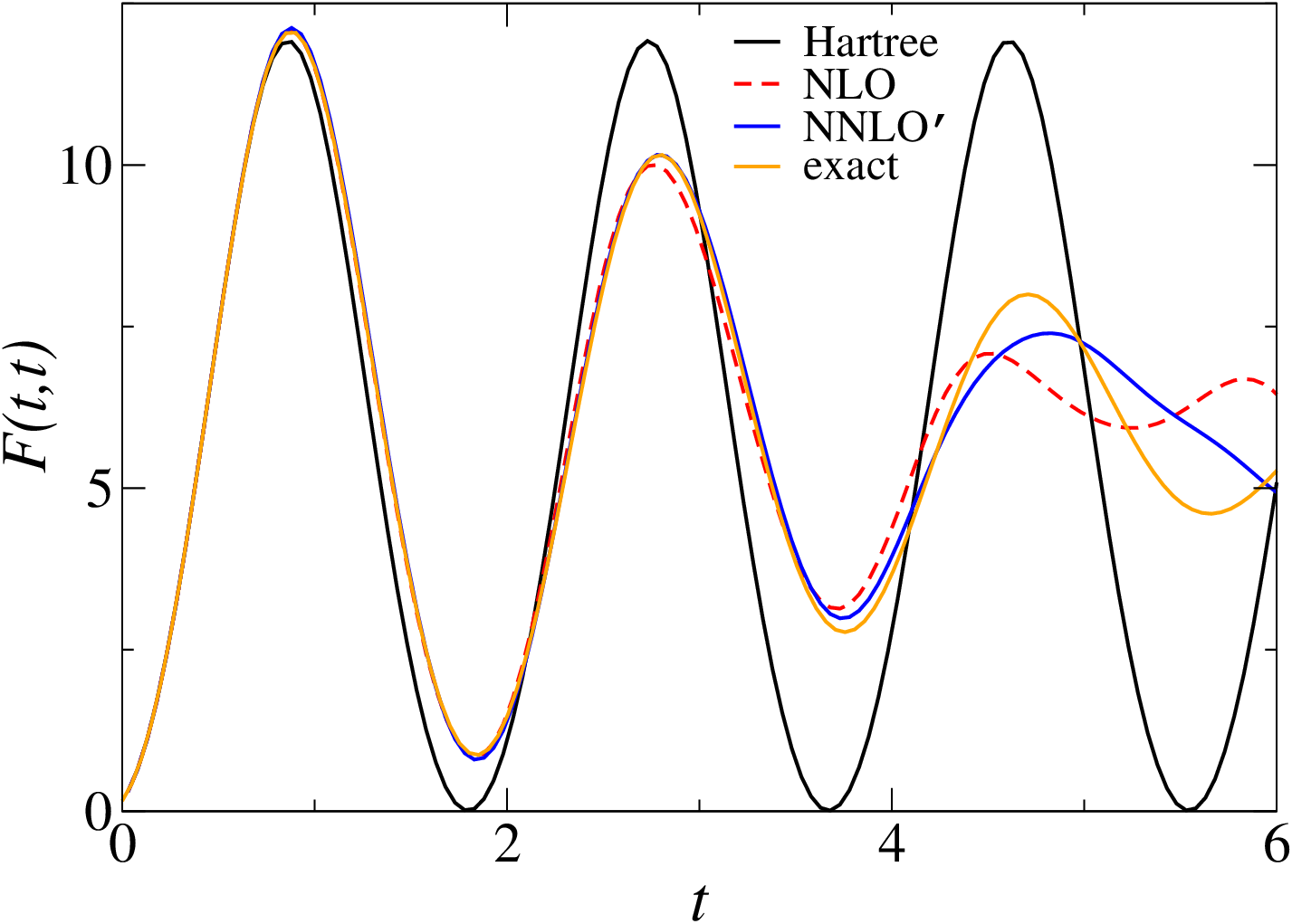,width=10cm}}
\caption{Time evolution of the equal-time correlator $F(t,t)$ for various 
approximations and the exact result from a numerical solution of the 
Schr\"odinger equation ($N=8$, $\lambda=1$, $\eta=5$, 
$\xi=0.4$, $\sigma=1$). 
}
\label{figN8}
\vspace*{1cm}
\centerline{\psfig{figure=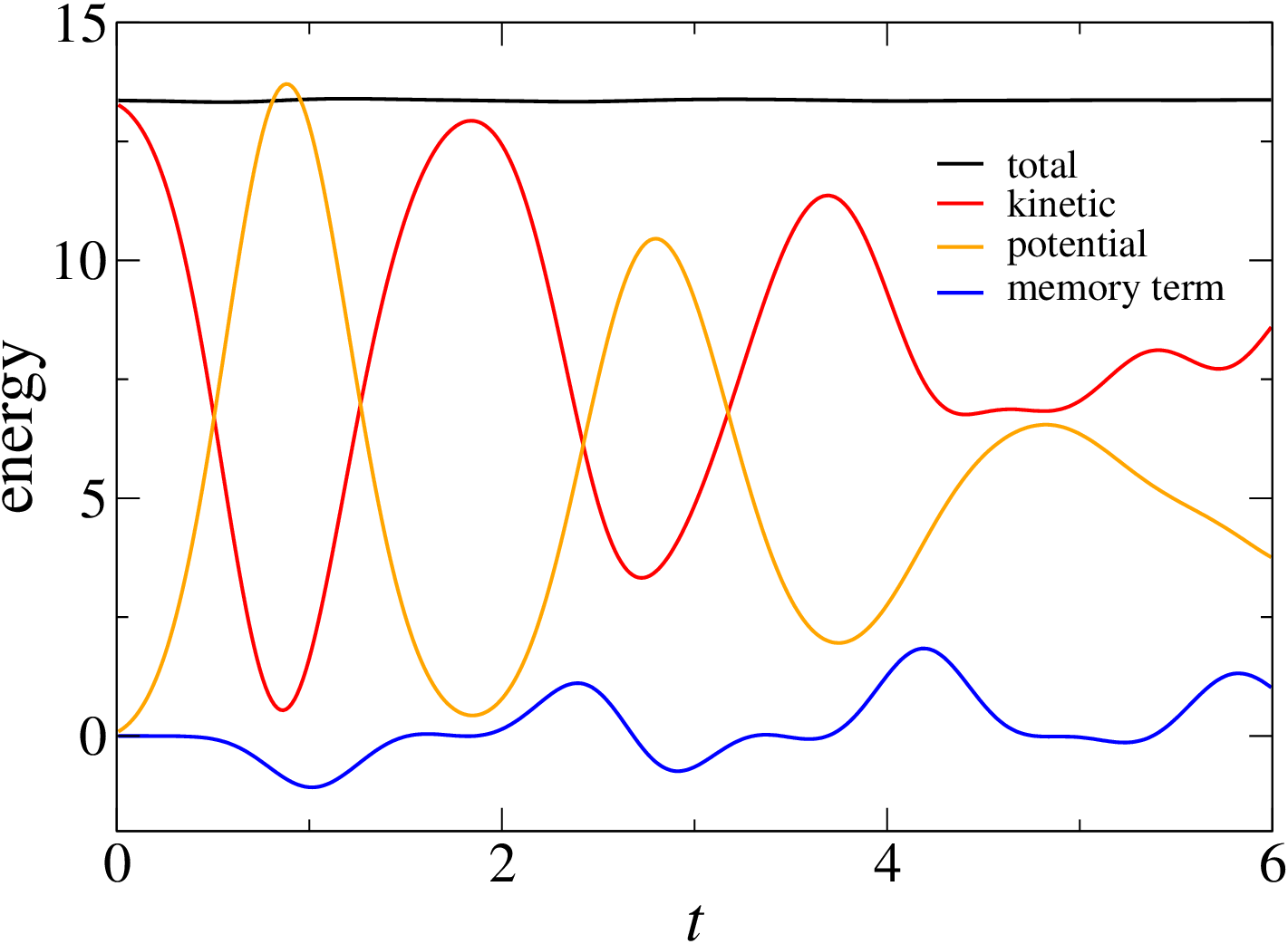,width=10cm}}
\caption{Time evolution of different energy components at 
N$^2$LO$^\prime$ (same parameters as in fig.\ \ref{figN8}).
 }
\label{NNLOenergy}
\end{figure}

 We now continue with a relative large value of $\eta=5$ and a smaller 
$\xi=0.4$ (corresponding to a squeezed initial state) since this yields 
an initially large amplitude and subsequent strong damping effects. For 
the sake of comparison with the numerical solution of the Schr\"odinger 
equation, we consider from now on pure states only, $\sigma=1$.

In fig.\ \ref{figN8} we show the evolution of the equal-time correlator 
$F(t,t)$ for the different levels of truncation, at $N=8$. In the Hartree 
approximation, all nonlocal terms on the RHS of the evolution equations 
(\ref{rhoeq}) are dropped and only the time-dependent mass parameter is 
preserved. Eqs.\ (\ref{Drhoeq}) are dropped altogether. At NLO the 
one-loop self energies (\ref{Prhoeq}) without internal vertices are kept. 
As explained at the end of the previous section, at N$^2$LO we only keep 
the two-loop self energies, with up to two internal vertices, in addition 
to the one-loop self energies. We will refer to this truncation as the 
N$^2$LO$^\prime$ approximation. Since it is derived from the two and 
three-loop diagrams in fig.\ \ref{figNLONNLO}, it is a 2PI self-consistent 
approximation.
 The result labelled with `exact' corresponds to the numerical solution of 
the Schr\"odinger equation, which is detailed in appendix \ref{appsch}.
  Energy conservation at N$^2$LO$^\prime$ is demonstrated in fig.\ 
\ref{NNLOenergy}, where the three components in eq.\ (\ref{eqenergy}) and 
their sum is shown. 

A close look at fig.\ \ref{figN8} shows that for early times both the NLO 
and the N$^2$LO$^\prime$ approximation are in quantitative agreement with 
the exact result. The amplitude in the Hartree approximation shows no sign 
of decreasing, due to a complete absence of dephasing in quantum 
mechanics.\footnote{This is different in field theory, where the 
amplitudes of equal-time correlation functions diminish in time due to 
dephasing. However, in the Hartree approximation the system still 
experiences recurrence. This shows that damping in the Gaussian 
approximation is not a result of irreversible loss of memory.} Around 
$t=2.7$, the NLO approximation starts to differ from the exact evolution, 
whereas the evolution at \NNLOp is capable to follow the exact evolution a 
bit longer. Around $t=4$ we find that both truncations fail to track the 
exact evolution and continue to evolve in an irregular fashion. We have 
verified that this behaviour is not due to the time discretisation. We 
also note that energy remains conserved. The irregular behaviour at later 
times seems to be peculiar to quantum mechanics and has been observed 
before in dynamics from truncated effective actions \cite{Mihaila:2000ib}. 
As far as we know it has not been observed in 2PI dynamics in the field 
theory case, where already the NLO approximation results in equilibration 
and thermalization. In this sense quantum mechanics, with only a finite 
number of degrees of freedom, is very different.

\begin{figure}
\centerline{\psfig{figure=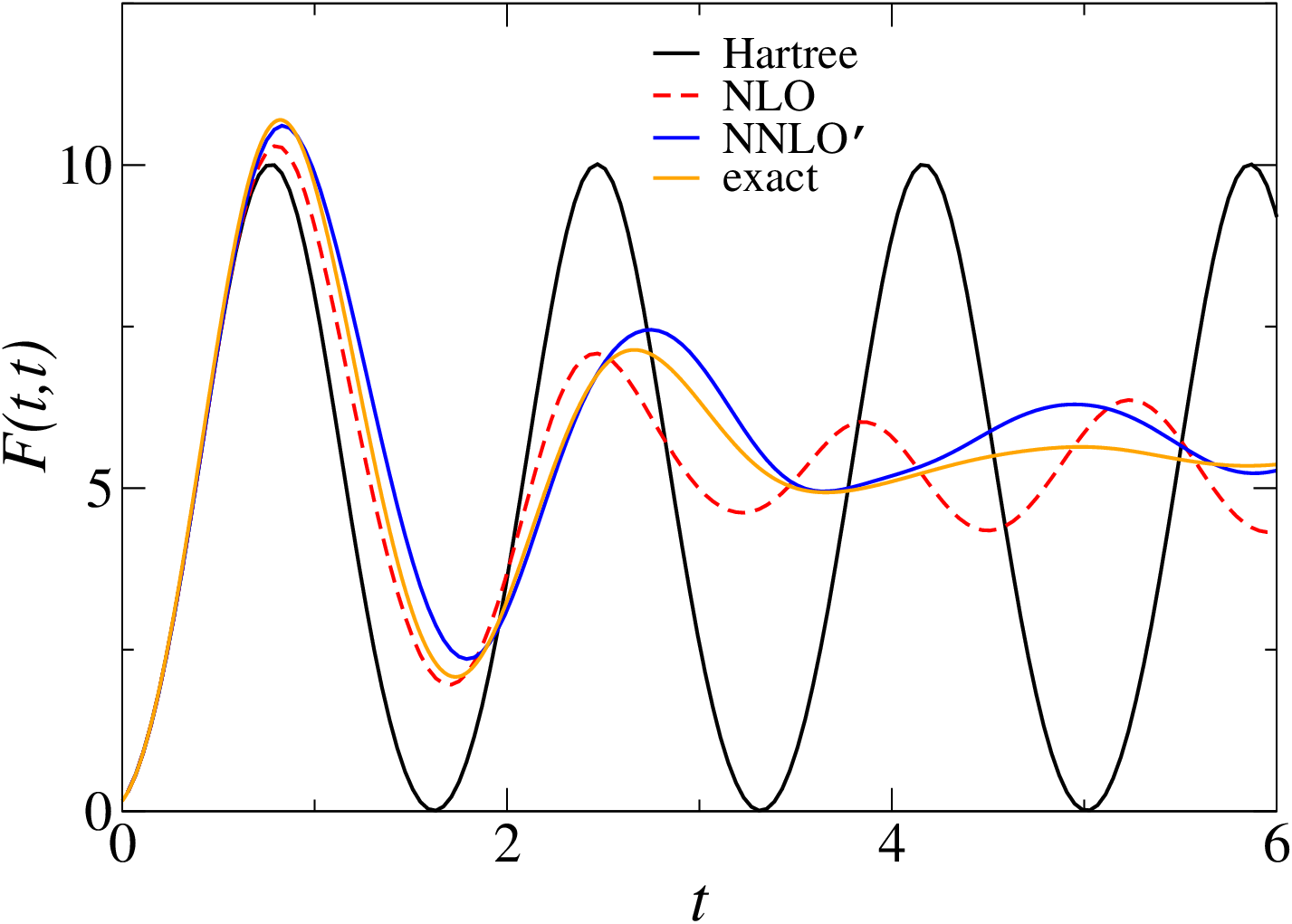,width=10cm}}
\caption{Same as Figure \ref{figN8}, for $N=2$.}
\label{compallqqz2}
\vspace*{1cm}
\centerline{\psfig{figure=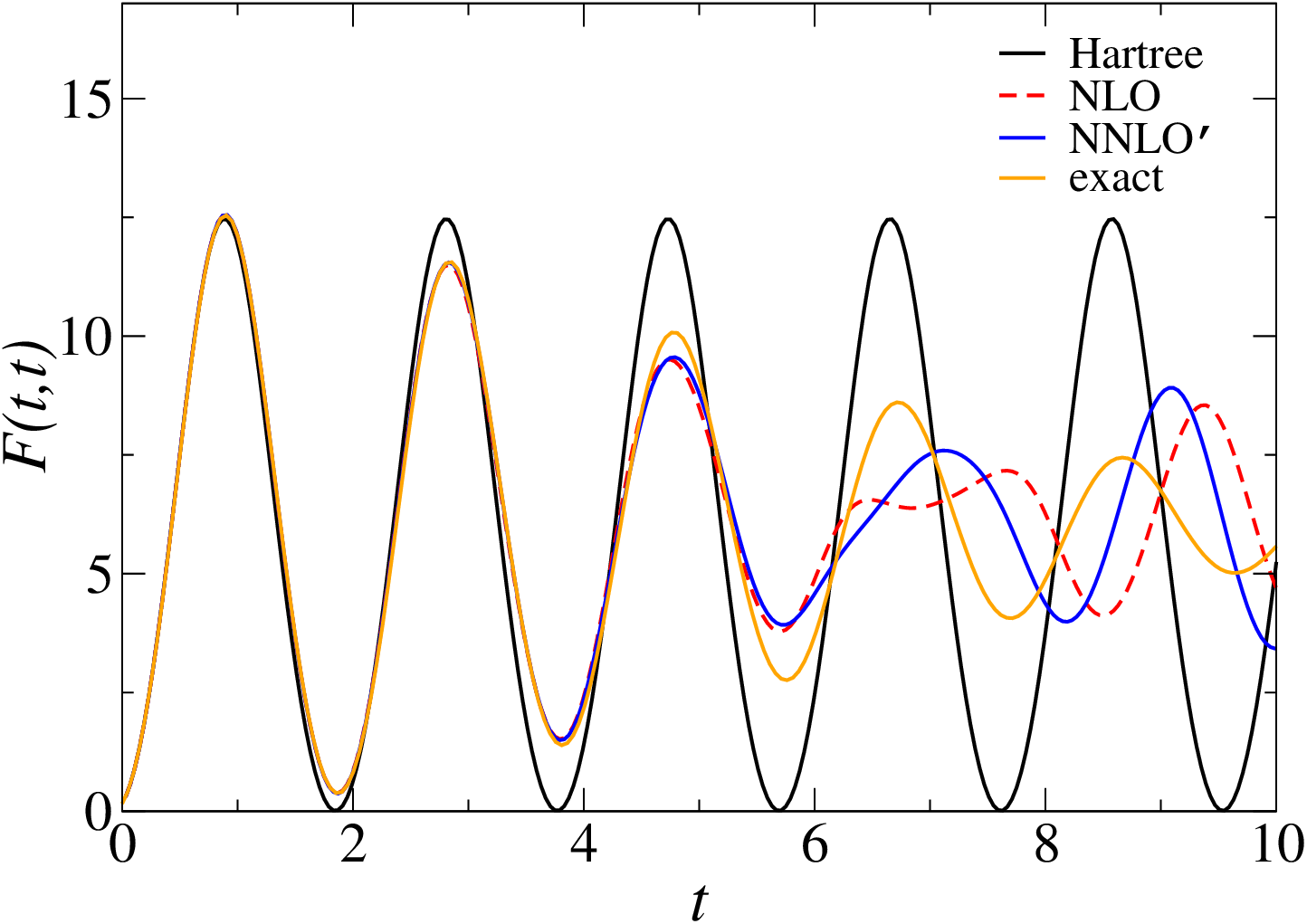,width=10cm}}
\caption{Same as Figure \ref{figN8}, for $N=20$.}
\label{compallqqz20}
\end{figure}

If we continue with a comparison at early times, we note that for $N=8$ 
the \NNLOp approximation works slightly better than the NLO approximation. 
This can be investigated further by looking at different values for $N$. 
Since the \NNLOp term is suppressed by $1/N$, it is expected that the 
difference between the \NNLOp and the NLO evolution will be largest for 
smaller $N$, while for larger $N$ the $1/N$ expansion itself is better 
behaved and the difference between \NNLOp and NLO is reduced.

This qualitative picture is confirmed by first going to smaller $N$. In 
fig.\ \ref{compallqqz2} we show again the equal-time correlation function, 
but now for $N=2$. As expected, the evolution ceases to follow the exact 
one earlier, but a close look at the first maximum indicates that it is 
first the Hartree approximation that breaks down, subsequently the NLO 
approximation and finally the \NNLOp approximation. At later times 
irregular behaviour is again observed. We find therefore that at early 
times an increase in the order of the truncation has a quantitatively 
correct effect. In fig.\ \ref{compallqqz20} the equal-time correlation 
function is shown again, but now for larger $N=20$. In this case the 
effect of adding the \NNLOp contribution is much less important, as 
expected, and we find that the \NNLOp evolution follows NLO rather than 
the exact curve, consistent with an effective convergence of the expansion 
for large $N$. Both curves follow the exact evolution for longer than in 
the cases shown above.

\begin{figure}
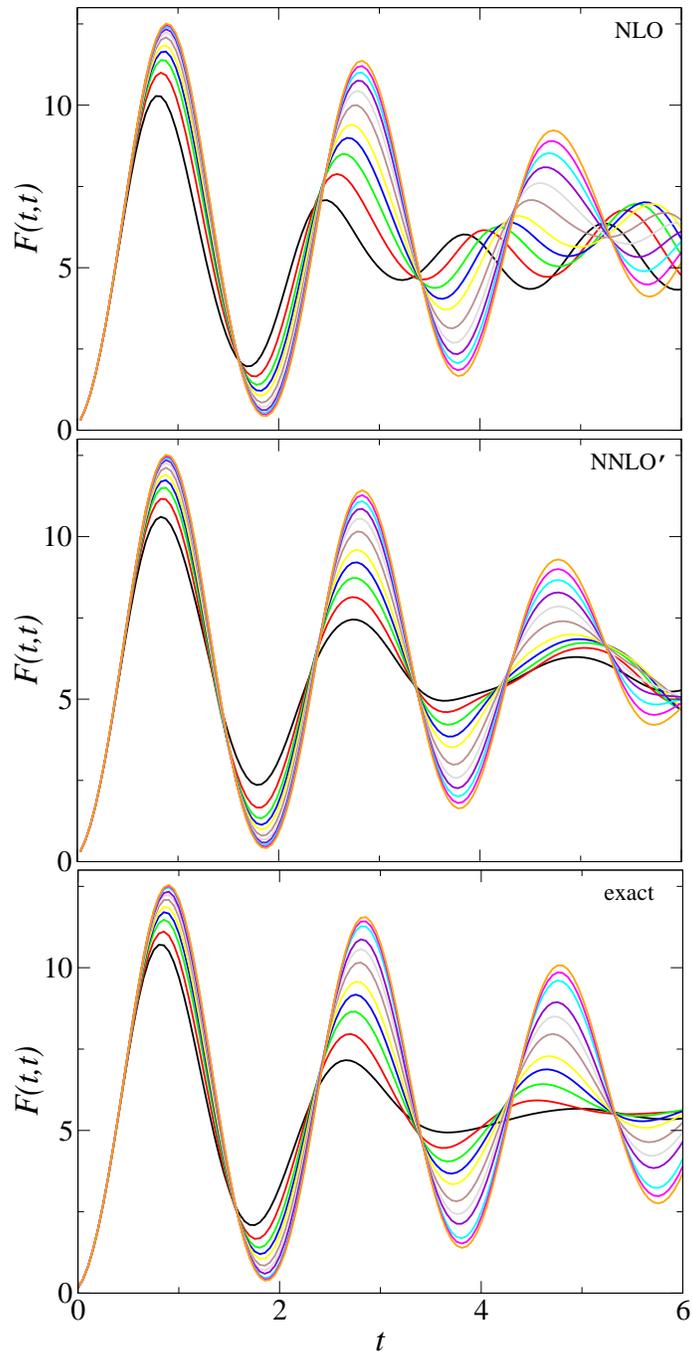

\centerline{\psfig{figure=comp_NLO_l1_allN_qq_2G.eps,width=8.9cm}}
\vspace*{-0.1cm}
\centerline{\psfig{figure=comp_NNLO_l1_allN_qq_2G.eps,width=8.9cm}}
\vspace*{-0.1cm}
\centerline{\hspace*{-0.015cm}
 \psfig{figure=comp_exact_l1_allN_qq_G.eps,width=9cm}}
\vspace*{-0.2cm}
\caption{Time evolution of $F(t,t)$ for NLO (upper), N$^2$LO$^\prime$ 
(middle) and exact (lower)  dynamics for various $2\leq N\leq 20$ 
(other parameters as in fig.\ \ref{figN8}).}
\label{NNLOallN}
\end{figure}

 \begin{figure}
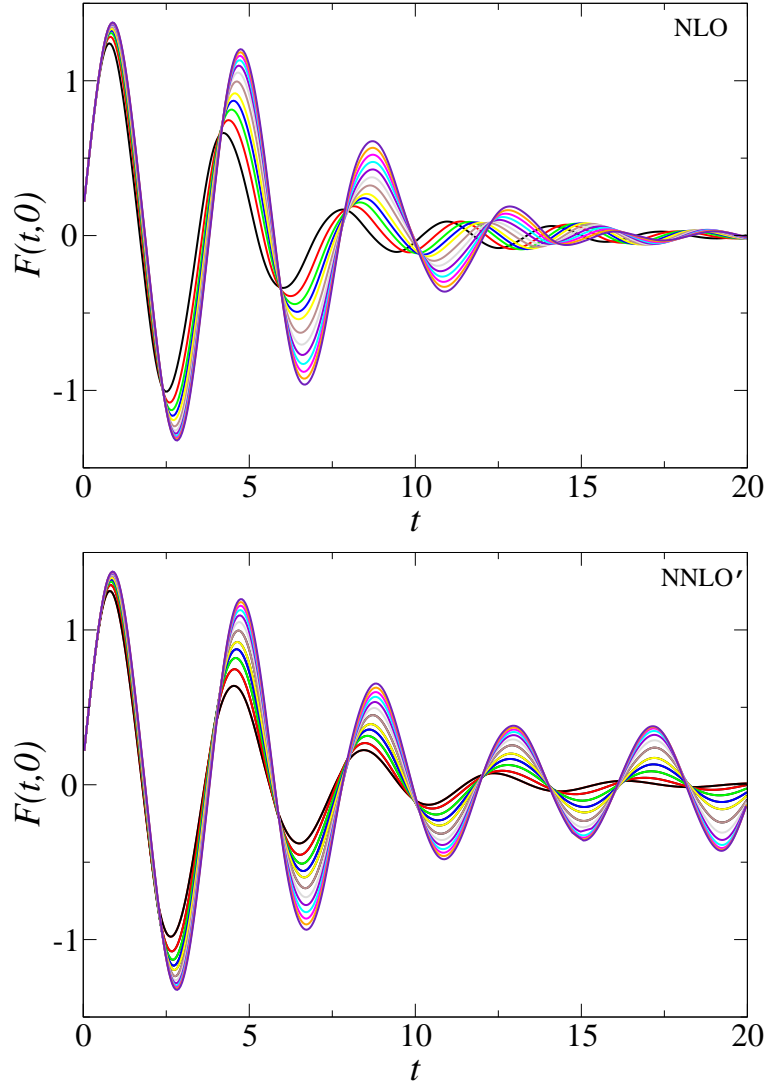
 
 \centerline{\psfig{figure=comp_NLO_l1_allN_F0t_G.eps,width=10cm}} 
 \vspace*{0.2cm}
 \centerline{\psfig{figure=comp_NNLO_l1_allN_F0t_G.eps,width=10cm}}  
 \caption{Unequal-time correlation 
 function $F(t,0)$ for NLO (upper) and N$^2$LO$^\prime$ (lower) dynamics
  for various $2\leq N\leq 20$ 
  (other parameters as in fig.\ \ref{figN8}).
} 
 \label{memoryloss} 
 \end{figure} 

A comparison between the NLO, the \NNLOp and the exact evolution is shown 
in figure \ref{NNLOallN}, for a wide range of $N$, from $N=2$ to $N=20$. 
It is observed that the exact solution appears to be intermediate between 
the NLO and \NNLOp evolution, with the evolution at \NNLOp performing 
slightly better, in particular in terms of the oscillation frequency at 
smaller $N$.

As mentioned above, thermalization and equilibration cannot be 
investigated in quantum mechanics. However, in analogy with field theory 
effective loss of memory can be studied. 
 In fig.\ \ref{memoryloss} we show the unequal-time two-point function 
$F(t,0)$ for various values of $N$. Because it is nonlocal in time, this 
correlator is not immediately accessible from the solution of the 
Schr\"odinger equation. Smaller values of $N$ correspond to a more rapid 
decrease of the amplitude, as expected. Maybe surprisingly, the memory 
appears to be washed out faster at NLO than at \NNLOp. However, this 
conclusion should be treated with care, since the approximations fail to 
track the exact evolution after some ($N$ dependent) time, as was shown 
above for the equal-time correlation functions.

\section{Outlook} 
\label{outlook}
\setcounter{equation}{0} 

We considered nonequilibrium dynamics in the $O(N)$ model, employing the 
2PI-$1/N$ expansion to N$^2$LO. We presented the explicit expressions for 
the three-loop approximation in the auxiliary-field formalism, to which we 
refer as N$^2$LO$^\prime$, and indicated how to obtain to full N$^2$LO 
contribution. The resulting evolution equations were solved numerically 
for quantum mechanics.

While the qualitative change in the nonequilibrium evolution when going 
from Gaussian (LO) approximations to NLO is enormous, the impact when 
changing from NLO to N$^2$LO is reassuringly small. This indicates that 
the 2PI-$1/N$ expansion is effectively rapidly converging and that higher 
order effects give quantitative corrections only.
 We found that at early times the evolution at \NNLOp performs slightly 
better than at NLO. This is especially visible at small $N$, where the 
higher-order contribution is not suppressed. At large times we found that 
all truncations break down, but we believe that this is special for 
quantum mechanics since it has not been observed in field theory, where 
already the NLO approximation has been seen to perform very well. It would 
therefore be very interesting to implement the N$^2$LO truncation, or at 
least the three-loop diagram as we considered here, in a $1+1$ dimensional 
scalar $O(N)$ model and test the apparent convergence quantitatively in 
field theory, also for late times.
  Alternatively, with some effort one may also be able to do the full 
N$^2$LO approximation in the case of quantum mechanics.

Finally, in this paper we only considered the symmetric phase. However, an 
extension to the broken phase at N$^2$LO is straightforward, since it 
involves only one additional two-loop diagram in the effective action, 
yielding new self energy type contributions without internal vertices 
\cite{Aarts:2002dj}. In fact, this diagram has been included already in 
the so-called `bare vertex approximation' (BVA) \cite{Mihaila:2000sr}.

                                                                                
\vspace*{0.5cm}
\noindent
{\bf Acknowledgments.}
 G.A.\ is supported by a PPARC Advanced Fellowship. A.T. is supported by 
the PPARC SPG ``Classical Lattice Field Theory''.



\appendix
\renewcommand{\theequation}{\Alph{section}.\arabic{equation}}


\section{Loop expansion \label{apploopexpansion}}
\setcounter{equation}{0}

\begin{figure}[b]
\centerline{\psfig{figure=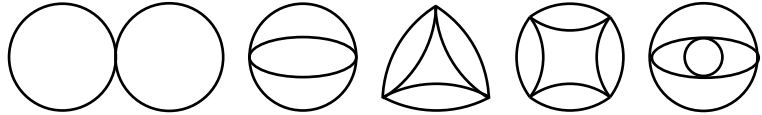,height=1.5cm}}
\caption{Loop expansion: contribution to $\Gamma_2$ up to fifth loop order. 
}
\label{figloop}
\end{figure}

In order to make a connection between the $1/N$ expansion and the ordinary 
loop expansion (see fig.\ \ref{figloop}), we give here the expression up 
to five loops in the loop expansion in the $O(N)$-symmetric case. We write 
the 2PI contribution as $\Gamma_2 = \sum_{l=2}^{\infty} \Gamma_2^{(l)}$ 
and find to fourth order
 \bea
\Gamma_2^{(2)} &=& -\frac{\lambda}{8}\frac{(N+2)}{3}\int_x G^2(x,x), \\
\Gamma_2^{(3)} &=& \frac{i\lambda^2}{48}\frac{(N+2)}{3N}\int_{xy} 
G^4(x,y),\\
\Gamma_2^{(4)} &=& \frac{\lambda^3}{48}\frac{(N+2)(N+8)}{27N^2}\int_{xyz} 
G^2(x,y)G^2(x,z)G^2(z,y).
\eea
At fifth order two diagrams contribute, $\Gamma_2^{(5)} = \Gamma_2^{(5a)}
+ \Gamma_2^{(5b)}$, which read
\bea
\nn
\Gamma_2^{(5a)} &=& 
-\frac{i\lambda^4}{128}\frac{(N+2)(N^2+6N+20)}{81N^3} \times\\
&&
\int_{xyzw} 
G^2(x,y)G^2(y,z)G^2(z,w)G^2(w,x)\\
\nn
\Gamma_2^{(5b)} &=& 
-\frac{i\lambda^4}{32}\frac{(N+2)(5N+22)}{81N^3} \times\\
\label{eqeye}
&&\int_{xyzw} 
G^2(x,y)G(x,z)G(x,w)G^2(z,w)G(y,z)G(y,w).
\eea
 In the 2PI-$1/N$ expansion to N$^2$LO, $\Gamma_2^{(2)}$ and 
$\Gamma_2^{(3)}$ are completely included, while from $\Gamma_2^{(4)}$ and 
$\Gamma_2^{(5a)}$ the NLO and N$^2$LO parts are taken into account. The 
eye diagram $\Gamma_2^{(5b)}$ starts at N$^2$LO only. Part of the leading 
N$^2$LO contribution is taken into account via the three-loop diagram in 
the 
auxiliary-field formalism and part via the four-loop diagram. 
The eye diagram is special since it is the first diagram that contributes 
to the bulk viscosity in the weak coupling limit 
\cite{Jeon:1994if,Calzetta:1999ps}.


\section{Contour integrals \label{appcontourintegrals}}
\setcounter{equation}{0}

\begin{figure}[t]
\centerline{\psfig{figure=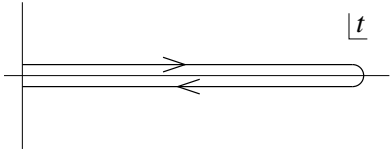,height=2.cm}}
\caption{Keldysh contour in the complex-time plane.
}
\label{figcontour}
\end{figure}

The evolution equations are formulated along the Keldysh contour in the 
complex-time plane, see fig.\ \ref{figcontour}. Splitting the propagators 
and self energies in statistical and spectral components yields the real 
and causal equations discussed in section \ref{causalequations}. While for 
the NLO approximation this procedure is straightforward, it becomes 
cumbersome when multi-loop contour integrals are encountered and integrals 
over products of sgn-functions have to be evaluated. Here we give some 
general expressions we find useful.

First consider integrals over products of $\Theta$-functions. Since the 
contribution from the upper and lower part of the contour differ only in 
sign, one finds e.g.\ that
\be
\int_\C dz^0 =  0, 
\;\;\;\;\;\;\;\;\;\;\;\;\;\;\;\;
\int_\C dz^0\, \Theta_\C(x,z) = \int_0^{x^0} dz^0,
\ee
or in general
\be
\int_\C dz^0 \prod_{i=1}^M \Theta_\C (x_i^0,z^0) = 
\sum_{i=1}^M \left[ \prod_{j\neq i}\Theta_\C (x_j^0,x_i^0)\right] 
\int_0^{x_i^0}dz^0.
\ee
Here we have taken the initial time at $z^0=0$.  
These results can be employed for integrals over products of 
sgn-functions, using that $\sgn_\C(x,y) = 2\Theta_\C(x,y) - 1$, yielding 
e.g.\
\be 
 \int_\C dz^0\,\sgn_\C(x,z) = 2 \int_0^{x^0} dz^0,
\ee
or in general
\be
\label{eqB4}
\int_\C dz^0 \prod_{i=1}^M \sgn_\C (x_i^0,z^0) = 
2\sum_{i=1}^M \left[ \prod_{j\neq i}\sgn_\C (x_j^0,x_i^0)\right] 
\int_0^{x_i^0}dz^0.
\ee
In a theory with quartic interactions, (\ref{eqB4}) is needed with $M\leq 
4$, for which we have verified this identity explicitly.

\section{Numerical solution of the Schr\"odinger equation}
\label{appsch}
\setcounter{equation}{0}

In the case of quantum mechanics, we have the option to solve the 
Schr\"odinger equation numerically for the full wavefunction. This only 
applies to pure states, $\sigma=1$. The system corresponds to a 
spherically symmetric anharmonic oscillator in $N$ dimensions, and by 
imposing $O(N)$ symmetry and make suitable redefinitions we can reduce the 
problem to that of a particle on a halfline in one dimension 
\cite{Mihaila:2000ib}.

If the wave function is written as the product of a radial function, 
depending on the (rescaled) radial coordinate $0< r=\sqrt{q_aq_a/N}< 
\infty$, and hyperspherical harmonics, depending on the $N-1$ angles, 
\be
\Psi(q_a,t) = \left(Nr^2\right)^{(N-1)/4} \Phi(r,t) Y(\Omega),
\ee
the $O(N)$ symmetric problem is reduced to the effective 
one-dimensional Schr\"odinger equation \cite{Mihaila:2000ib}
\be
 \label{eqsch}
 \frac{i}{N}\frac{\partial}{\partial t} \Phi(r,t) = {\cal H}_{\rm eff} 
\Phi(r,t),
\ee
with the Hamiltonian
\be
 {\cal H}_{\rm eff}  = 
-\frac{1}{2N^2} \frac{\partial^2}{\partial r^2} + U(r),
\ee
and the effective potential
\be
 U(r) = \frac{(1-1/N)(1-3/N)}{8 r^2} + \half m^2 r^2 + 
\frac{\lambda}{4!}r^4.
\ee
 For a pure state the initial radial wave function, corresponding to the 
density matrix discussed in section \ref{quantummechanics}, is
 \be
 \Phi(r,0) =  \frac{1}{\sqrt{\half\Gamma(N/2)}}
\left(\frac{N}{2\xi^2}\right)^{N/4}
r^{(N-1)/2}
\exp\left[ - N \left( \frac{1}{4\xi^2} + i\frac{\eta}{2\xi}\right) r^2
\right].
\ee
This initial wave function is normalized with respect to the innerproduct
\be
 \bra \Phi|\Phi\ket =  \int_0^\infty dr\, \Phi^\dagger(r,t) \Phi(r,t).
\ee
 We solve eq.\ (\ref{eqsch}) numerically using the second-order 
Crank-Nicholson differencing scheme \cite{NumRec}. For the time intervals 
shown, both unitarity and energy are preserved better than 1 in $10^{12}$.


\end{document}